\documentclass[conference]{IEEEtran}
\ifCLASSINFOpdf
\else
\fi

\usepackage{amssymb}
\usepackage{setspace}
\usepackage{graphicx}
\usepackage[tight,footnotesize]{subfigure}

\begin{document}

\title{EPDA: Enhancing Privacy-Preserving Data Authentication for Mobile Crowd Sensing}

\author{\IEEEauthorblockN{Jingwei Liu\IEEEauthorrefmark{1},
Fanghui Cai\IEEEauthorrefmark{1},
Longfei Wu\IEEEauthorrefmark{2},
Rong Sun\IEEEauthorrefmark{1},
Liehuang Zhu\IEEEauthorrefmark{4} and
Xiaojiang Du\IEEEauthorrefmark{3}}
\IEEEauthorblockA{\IEEEauthorrefmark{1}State Key Lab of ISN, Xidian University, Xi'an, 710071, China.\\ Email: \{jwliu, rsun\}@mail.xidian.edu.cn, caidoreen@163.com}
\IEEEauthorblockA{\IEEEauthorrefmark{2}Department of Mathematics and Computer Science, Fayetteville State University, Fayetteville, NC 28301, USA\\ Email: lwu@uncfsu.edu}
\IEEEauthorblockA{\IEEEauthorrefmark{3}Department of Computer and Information Sciences, Temple University, Philadelphia, PA 19122, USA.\\ Email: dxj@ieee.org}
\IEEEauthorblockA{\IEEEauthorrefmark{4}School of Computer Science, Beijing Institute of technology, Beijing 100081£¬China\\ Email: liehuangz@bit.edu.cn}
}

\maketitle

\begin{abstract}
As a popular application, mobile crowd sensing systems aim at providing more convenient service via the swarm intelligence. With the popularity of sensor-embedded smart phones and intelligent wearable devices, mobile crowd sensing is becoming an efficient way to obtain various types of sensing data from individuals, which will make people's life more convenient. However, mobile crowd sensing systems today are facing a critical challenge, namely the privacy leakage of the sensitive information and valuable data, which can raise grave concerns among the participants. To address this issue, we propose an enhanced secure certificateless privacy-preserving verifiable data authentication scheme for mobile crowd sensing, named EPDA. The proposed scheme provides unconditional anonymous data authentication service for mobile crowd sensing, by deploying an improved certificateless ring signature as the cryptogram essential, in which the big sensing data should be signed by one of legitimate members in a specific group and could be verified without exposing the actual identity of the participant. The formal security proof demonstrates that EPDA is secure against existential forgery under adaptive chosen message and identity attacks in random oracle model. Finally, extensive simulations are conducted. The results show that the proposed EPDA efficiently decreases computational cost and time consumption in the sensing data authentication process.
\end{abstract}
\IEEEpeerreviewmaketitle

\section{Introduction}
\begin{figure}[!t]
  \centering
  \includegraphics[width=8.5cm]{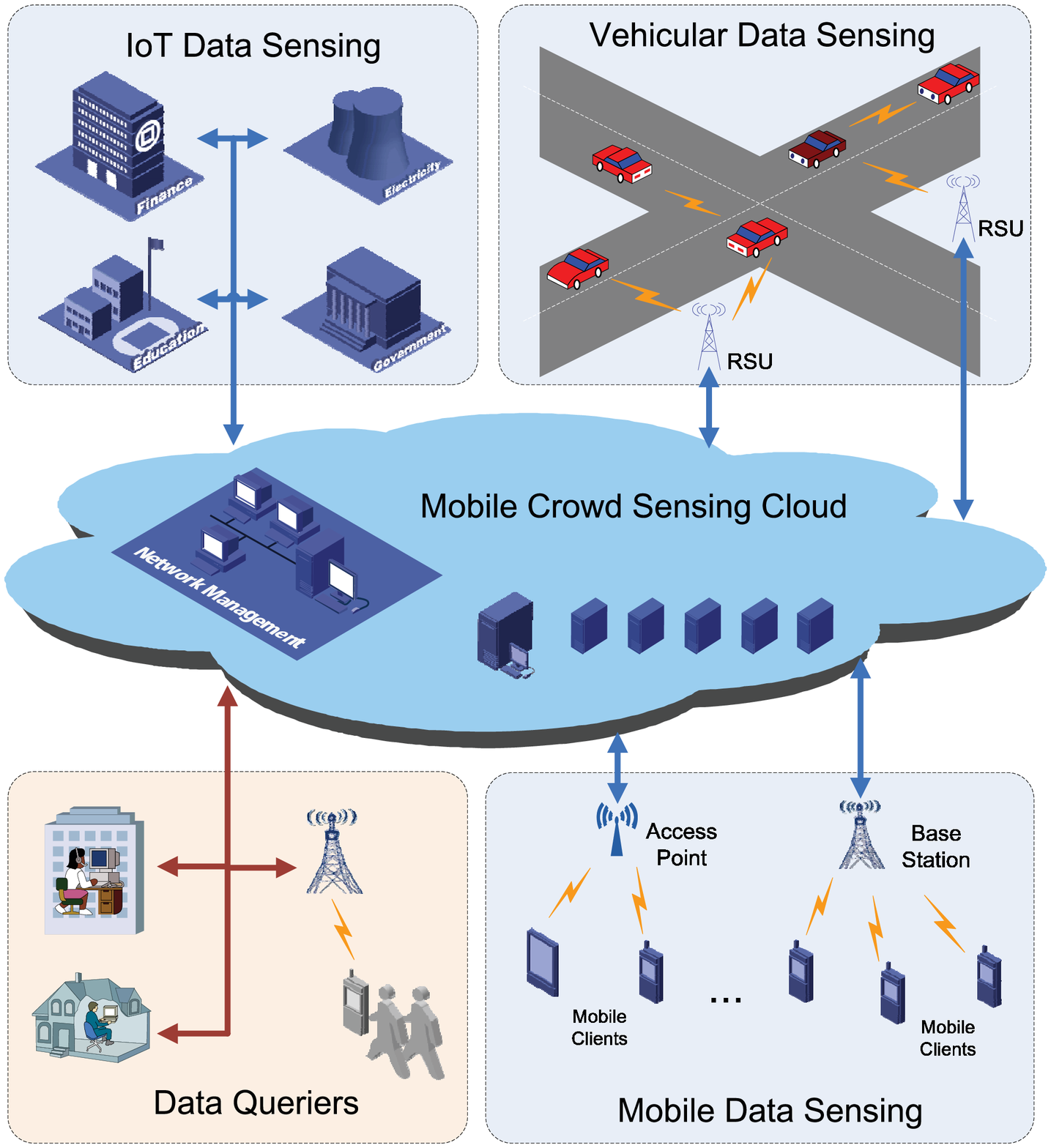}\\
  \caption{A simple architecture of MCS systems}
  \label{MCS_archi}
\end{figure}

In the big data era, a mass of mobile terminals (such as smart phones, tablets and laptops) equipped with a variety of sensors (e.g., GPS, accelerator, camera) are producing huge amount of sensing data. It changes the traditional crowd sensing mode to Mobile Crowd Sensing (MCS) \cite{RLPG12, GYL11, ZBCVZ14} as illustrated in Fig. \ref{MCS_archi}, where the sensing tasks could be released more quickly and conveniently, and the sensing data could be collected promptly. At present, MCS systems have been widely used in vehicular networks (for the location information and traffic data), body area networks (for physical bio-information) \cite{LZCK14, LHWSD17}, Internet of things (for the real-time condition), social networks and so on \cite{XCDG09, LD14}. This trend has accelerated the progress of smart cities.

Recently, more and more researchers have been studying the trend and providing a broad prospect of MCS systems, in which participants submit the sensing data or other requested information via their intelligent terminals to third parties who are interested in these data for specific purposes. The collected data may be very sensitive since it is likely to reveal the privacy of the device owner \cite{AIBB12, KFD10}, such as identity, location, health status, and personal activities. It may lead to many uncertain security threats and affect the enthusiasm of the participants. Therefore,  users' privacy-preserving and security issues should be taken into consideration seriously in MCS systems \cite{QM08, ZHQ12, DL05, DZNGC07, CWLHL14, LPC14, LJSG12, DGXC09, DC08, GZZF13, ZJGLL16}.

In \cite{LMLPCC10}, a typical architecture of MCS system was introduced, which usually includes three participants: a group of clients, a network management (NM), and a query service provider (SP). Upon receiving a sensing task, participants gather and submit the required sensing data to the cloud server of a SP. Once collecting enough sensing data, the SP forwards the result to the requester for further analysis. However, if MCS systems deploy traditional public key cryptography to authenticate these data, it may raise heavy burdens of the certificate verification and management. An available approach is to introduce Certificateless Public Key Cryptography (CL-PKC), which does not involve complicated certificate management any more. In 2003, Al Riyami and Paterson proposed the first certificateless public key cryptosystem (CL-PKC) \cite{AP03}. By combining the merits of traditional key management system (PKC) and identity-based cryptography (IDC) \cite{Sha84}, CL-PKC is used to implement the implicit certification (through users' IDs) to address the inherent key escrow problem in IDC (through the user's secret information). CL-PKC has been attracting more and more attentions in recent years \cite{GS05, HSMZ05, YHG06, ZWXF06, DLP08, HMSWW07, ZZ08}.

Ring signature is a kind of effective cryptogram essential \cite{RST01} to protect users' privacy, which was first introduced in 2001. Any member in a specific group can generate a signature anonymously on behalf of the group, and anyone else including the other members in the group can verify this signature. Since no information about the signer's actual identity is revealed, the verifier cannot determine which member generated the signature. However, designing a ring signature scheme based on certificateless cryptography is not trivial. In 2007, Chow et al. proposed the first certificateless ring signature (CL-RS) \cite{CY07}. After this original work, many certificateless ring signatures \cite{CWMZ09, ZZW07, Wan10} were published subsequently.

In this paper, by deploying an improved CL-RS, we proposed an enhanced certificateless privacy-preserving data authentication scheme for MCS systems. The proposed scheme is proved to be secure from existential forgery on adaptive chosen message and identity attacks in random oracle model, assuming that the k-Collision Attack Algorithm (k-CAA) problem and the Inverse Computational Diffie-Hellman (Inv-CDH) problem are intractable. Finally, the performance is evaluated. The simulation results show that the proposed EPDA is more efficient for the privacy-preserving MCS scenario.

The rest part of this paper is organized as follows. The preliminaries are introduced in section II. In Section III, the enhanced privacy-preserving data authentication scheme for MCS system is presented in detail, including the security analysis. In Section IV, the performance of EPDA is evaluated. Finally, the conclusion is given in Section V.

\begin{table*}
\begin{center}
\footnotesize
\caption{Notations}\label{tabl:notation}
\begin{tabular}{c|c||c|clllll}
\hline
 $q$ & a large prime number & $\textsf{h}(\cdot)$ & secure hash function $\textsf{h}(\cdot): {\{0, 1\}}^\ast\rightarrow \mathbb{Z}^*_q$ \\
 $G_1$ & a cyclic additive group of order $q$ & $P$ & a generator of $G_1$ \\
 $G_2$ & a cyclic multiplicative group of order $q$ & $g$ & a generator of $G_2$ \\
 $\mathcal{A}_{1,2}$ & two types of adversaries & $data$ & sensing data \\
 $\mathcal{C}$  &  a challenger  &  $\sigma$ & digital signature \\
 $P_{pub}$  &  network manager's public key  &  $s$ & network manager's private key \\
 $R_{ID_i}$  &  the public key of the user with $ID_i$  &  $S_{ID_i}$  &  the private key of the user with $ID_i$  \\
 $R$  &  the public key set of participants  &  $L$  &  the identity set of participants \\
 \hline
\end{tabular}
\end{center}
\end{table*}

\section{Preliminaries}
To facilitate the understanding of the cryptogram essential in EPDA, we introduce the basic definitions and the properties of bilinear pairings over elliptic curve group. Also, we give the security model for EPDA. For easier illustration, Table~\ref{tabl:notation} lists some important notations which will be further explained where they occur for the first time.

\subsection{Bilinear Pairings}
The bilinear pairings of algebraic curves are defined as a mapping: $G_1\times G_1\to G_2 $ where $(G_1, +)$ is a cyclic additive group generated by $P$, whose order is a prime $q$, and $(G_2, \cdot)$ is a cyclic multiplicative group of the same order $q$. Bilinear pairings have the following properties:
\begin{itemize}
\item
%\emph{Bilinear}:
Bilinear: $e(aU, bV) = e(U, V)^{ab}$, $\forall$ $U, V \in G_1$ and $a, b\in Z^*_q$. This can be related as $\forall$ $U, V, W\in G_1$, $e(U + V, W)=e(U, W)\cdot e(V, W)$ and $e(U, V + W)=e(U, V)\cdot e(U, W)$;
\item
Non-degenerate: There exists $U, V\in G_1$ such that $e(U, V)\not=g$, where $g$ denotes the identity element of $G_2$;
\item
Computable: There is an efficient algorithm to compute $e(U, V)$ for all $U,V\in G_1$.
\end{itemize}

To prove the security of EPDA, we assume the following hard problems in $G_1$:

\textbf{Definition 1.} \emph{k}-Collision Attack Algorithm Problem (\emph{k}-CAAP): Given a fixed and known integer $k$, a ($2k + 2$)-tuple $(t_1, \dots, t_k, P, Q = sP, {\frac{1}{t_{1}+s}}P, \dots, {\frac{1}{t_k + s}}P)\in Z_{q}^{k} \times G_{1}^{k+2}$, output a pair $(A, c)$ such that $A = \frac{1}{c+s}P$.

\textbf{Definition 2.} Inverse Computational Diffie-Hellman Problem (Inv-CDHP): Given $P$ and $aP$ for $a\in Z_{q}^{*}$, output $\frac{1}{a}P$.

\subsection{Security Model}
We assume there are two types of opponents with different capabilities in EPDA:

\begin{enumerate}
  \item $\mathcal{A}_I$ is an attacker who is able to replace public keys, extract partial private keys and make sign queries without the master secret key.
  \item $\mathcal{A}_{II}$ is an attacker who can obtain the master secret key. It may replace the public keys, extract partial private keys and make sign queries.
\end{enumerate}

We will prove the security properties of EPDA in the existential unforgeability under adaptive chosen message and identity attacks (EUF-CL-RS-CMIA2) model \cite{CY07} for both two types of adversaries. Also, we will give the analysis on anonymity. An opponent could reveal the real identity of any signer with the probability no more than $1/n$, while the member in the group is with the probability no more than $1/(n-1)$. Here, $n$ is the number of the group.

\section{Enhancing Privacy-Preserving Data Authentication for Mobile Crowd Sensing}
To meet the unconditional privacy-preserving demands in some certain MCS scenarios, we propose an enhanced privacy-preserving data authentication scheme for MCS system, which can preserve the anonymity of participants, by deploying an improved certificateless ring signature as the cryptogram essential.

\subsection{Design Objectives}
With different kinds of micro-sensors for location, temperature, and biomedical being integrated into the intelligent terminals, it has been possible for a mass of users to sense and upload sensing data to the MCS cloud upon different requests. For instance, an institution of public health service may intend to gather participants' bio-information, like heart rates, blood pressure and so on at different times to study the changing trend of these factors in a day to reveal the relationship of each other. Moreover, the transportation management bureau may make use of the sensing data for monitoring and managing the urban traffic situation. Crowd sensing data collected by various intelligent terminals bring a various of convenient services for the querying clients or institutions, as the ubiquitous access of the Internet enables nearly real-time feedback. It saves lots of time and cost for the sensing data requester. However, no matter how promising the MCS is, it will not be well accepted only when the principle privacy issue is resolved perfectly. For example, a user's sensing data might involve private information like identity, location, and so on. Leaking out these private information to the cloud servers or other users could cause critical privacy disclosure or even physical attacks \cite{CWLHL14, LPC14, DC08, GZZF13, ZJGLL16}. Therefore, the participants might not be willing to accept the sensing tasks on account of privacy issues.

To begin with, we assume there is a TTP (Trusted Third Party) in the MCS system defined as NM which can generate and certify cryptographic keys. All participants should interact with NM in advance for key distribution. In addition, the MCS is operating over insecure networks. Therefore, anonymity is a basic property and the existence of active rivals who attempt to subvert the real identities of MCS clients can not be ignored. Based on the above assumptions and considering the characteristic of the mobile crowd sensing system. We design an authentication scheme with the following properties:

\begin{itemize}
\item Achieving anonymous sensing data authentication regardless of particular MCS scenarios over insecure channels.
\item Operating with relatively low computational cost.
\end{itemize}

\subsection{Design Architecture}

\begin{figure}[htb]
  \centering
  \includegraphics[width=7cm]{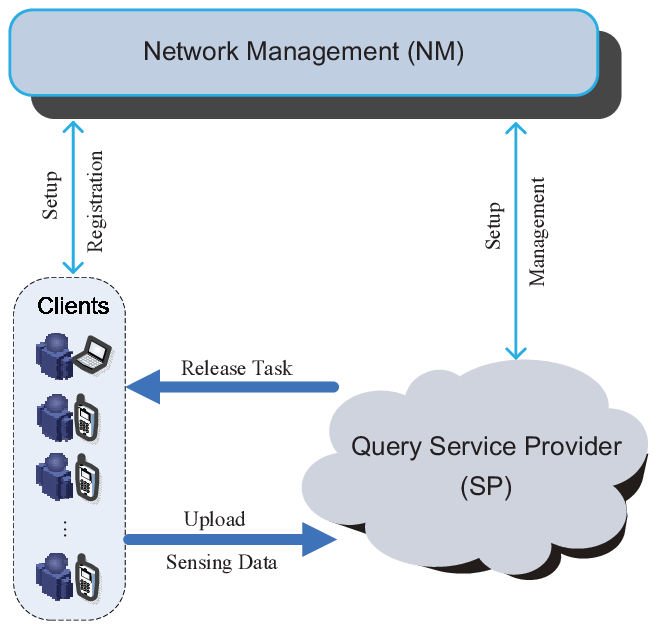}\\
  \caption{The framework of EPDA for MCS}\label{Framework_of EPDA}
\end{figure}

There will be three types of entities: MCS clients, the Network Manager (NM), and the Query Service Provider (SP) in the framework of EPDA, as shown in Fig. \ref{Framework_of EPDA}. In general, MCS clients refer to the participants in different regions equipped with smart phones to collect and submit various sensing data. SPs could be cloud servers of health organizations or research institutions. Additionally, NM is in charge of generating the partial private key for each user and publish identity indexes based on clients' public keys, which are used to authenticate all sensing data. In EPDA, NM is modeled as a trusted but curious third party. Note that the partial private key generated by NM is not sufficient to impersonate a legitimate client.

\subsection{The Enhanced Privacy-Preserving Data Authentication Scheme}
In general, the NM generates the system parameters firstly. Then, the NM generates the public keys and partial private keys for clients based on their identities in the register stage. Meanwhile, each client in MCS system calculates his/her partial private key based on a secure random number. The MCS participants use the private key for signing the sensing data, and SPs use a list of public keys for verifying respectively. When a client submits the signed sensing data to an SP, the SP will verify if the received sensing data is from a legitimate participant by checking the signature. If the verification equation holds, the uploaded data is valid. In EPDA, we will utilize an improved variant of CL-RS, which can ensure that though SPs can verify the signed sensing data in authentication procedure, they are not able to recover the actual identity of any participant. Supposing that SPs and clients are time synchronization, our protocol can be illustrated as follows:

\begin{figure}[tb]
  \centering
  \includegraphics[width=8.6cm]{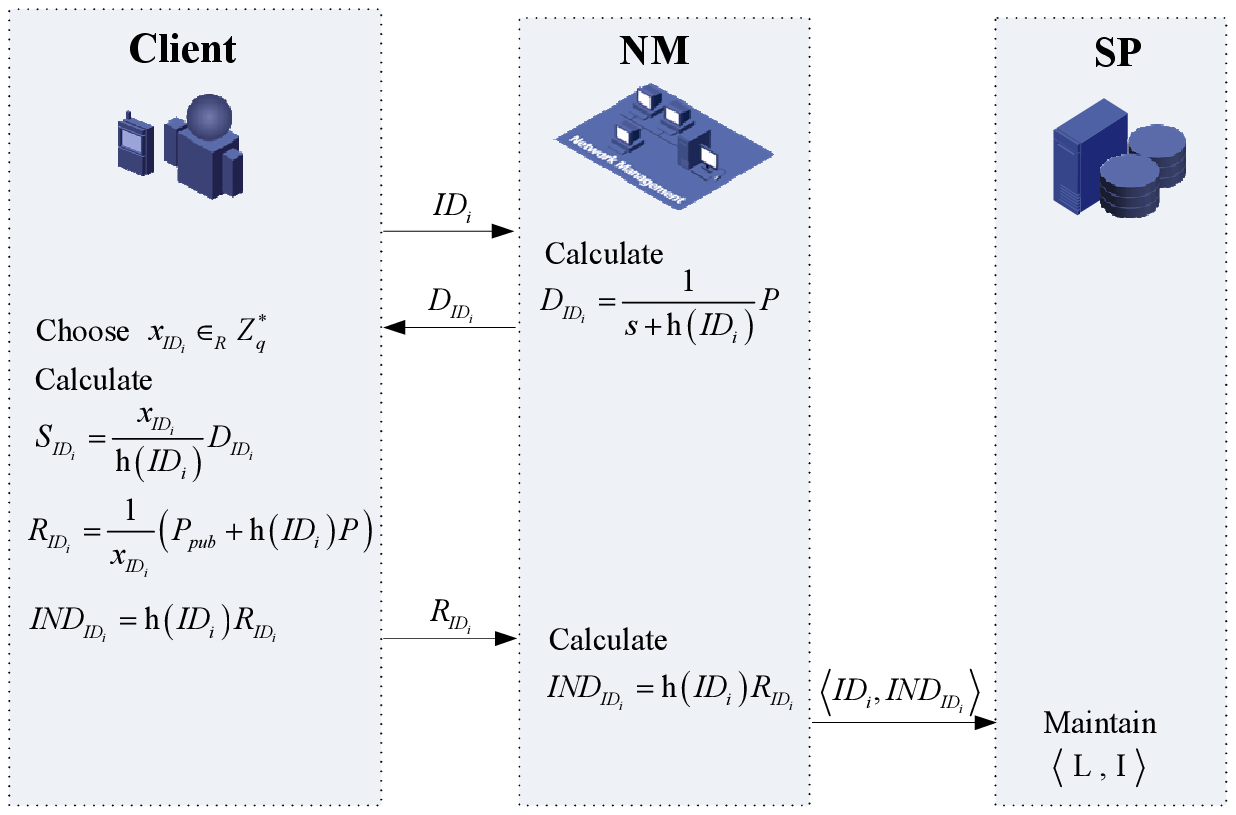}\\
  \caption{The registration of EPDA}\label{Registration}
\end{figure}

\begin{figure}[tb]
  \centering
  \includegraphics[width=8.2cm]{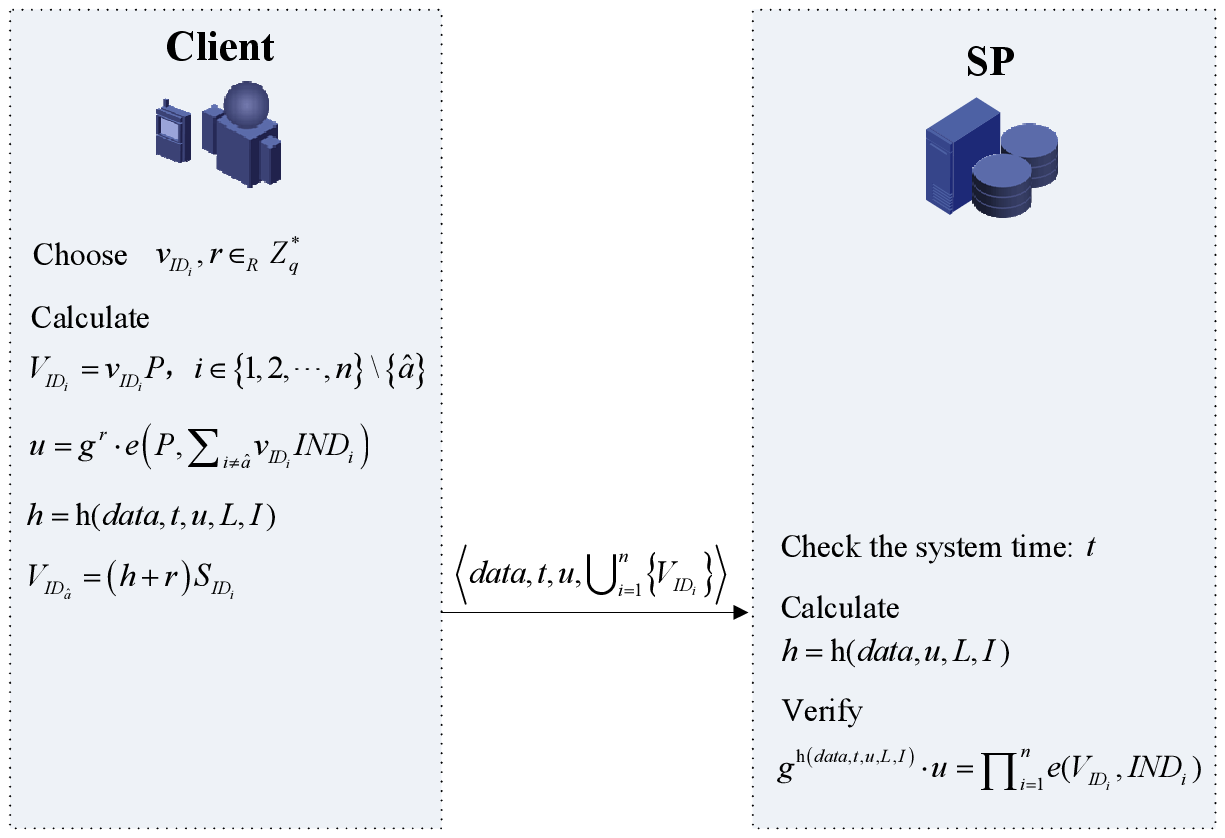}\\
  \caption{The flowchart of data authentication}\label{Authentication}
\end{figure}

\begin{itemize}
\item[1.] \emph{Initialization.} In the first place, given security system parameter $l$, NM generates keys for all participants in EPDA and initializes the authentication procedure. Let $(G_1, +)$ and $(G_2, \cdot)$ denote two cyclic groups of prime order $q>2^l$ and $e: G_1 \times G_1\rightarrow G_2$ be a pairing operator that satisfies the properties of bilinear and nondegenerate. Let $L = \{ID_1, ID_2, \cdots, ID_{n}\}$ denote the set of identities of $n$ clients and $R = \{R_{ID_1}, R_{ID_2}, \cdots, R_{ID_n}\}$ be the set of corresponding public keys. The NM determines its public or private key pair $\langle P_{pub}, s \rangle$, where $P_{pub}=sP$, and publicizes the system parameters $\langle l, G_1, G_2, q, P, e, \textsf{h}, Q_{NM} \rangle$.

\item[2.] \emph{Registration.} To accomplish MCS tasks distributed by any SP, a client need register to the NM. The whole registration steps shown in Fig. \ref{Registration} should be performed in turn:
\begin{itemize}
\item[a.]The client sends his/her $ID_i$ to the NM firstly.

\item[b.]Upon receiving $ID_i$, the NM calculates $D_{ID_i} = \frac{1}{s+q_{ID_i}}P$, where $q_{ID_i}=\textsf{h}(ID_i)$, and sends $D_{ID_i}$ back to the client.

\item[c.]The client chooses a random $x_{ID_i}\in Z_{q}^{*}$, and computes $S_{ID_i} = \frac{x_{ID_i}}{q_{ID_i}}D_{ID_i}$, $R_{ID_i} = \frac{1}{x_{ID_i}}(P_{pub} + q_{ID_i}P)$ and $IND_{ID_i}=q_{ID_i}R_{ID_i}$. Then s/he sends $R_{ID}$ to the NM.

\item[d.]The NM calculates $IND_{ID_i}=q_{ID_i}R_{ID_i}$ and adds a record of $\langle ID_i, R_{ID_i}, IND_{ID_i} \rangle$ to the database. Then, the NM sends $\langle ID_i, IND_{ID_i} \rangle$ to the SP. The SP maintains two lists: $L = \{ID_1, ID_2, \cdots, ID_n\}$ and $I = \{ IND_{ID_1}, IND_{ID_2}, \cdots, IND_{ID_n}\}$.
\end{itemize}

\item[3.] \emph{Uploading.} Upon receiving a sensing task from the SP, each participant gathers and uploads the required sensing data. Firstly, s/he chooses $v_{ID_i}, r\in Z_{q}^{*}$ randomly, and then computes $V_{ID_{i}} = v_{ID_i}P$, $i\in \{1, 2, \cdots, n\}\backslash\{\hat{a}\}$, $u = g^{r}\Pi_{i\neq \hat{a}}e(V_{ID_i}, IND_{ID_i}) = g^{r}e(P, \sum_{i\neq \hat{a}}v_{ID_{i}}IND_{ID_i})$, $h = \textsf{h}(data, t, u, L, I)$ and $V_{ID_{\hat{a}}} = (h+r)S_{ID_{\hat{a}}}$, where $t$ is the system time to keep the freshness of the messages. Eventually, the participant outputs the signature on sensing $data$ as $\sigma = \{u,\bigcup_{i = 1}^{n}\{V_{ID_{i}}\}\}$ and uploads it to the corresponding query service provider.

\item[4.] \emph{Verify.} As shown in Fig. \ref{Authentication}, the SP first checks the system time $t$, and then verifies the signature $\sigma = \{u,\bigcup_{i=1}^{n}\{V_{ID_{i}}\}\}$ on the submitted sensing $data$, by checking if $g^{\textsf{h}(data, t, u, L, I)}\cdot u = \prod^n_{ i = 1 }e(V_{ID_{i}}, IND_{ID_i})$ holds. If it does, the SP accepts the $data$. Otherwise, the SP discards this submission.

\end{itemize}

\subsection{Security Analysis}

In this section, we give the security proof in EUF-CL-RS-CMIA2 model.

\noindent\textbf{\emph{Theorem 1.}} \emph{The proposed EPDA is existential unforgeable against both $\mathcal{A}_I$ and $\mathcal{A}_{II}$ adversaries in the random oracle model under intractability assumptions of k-CCAP and Inv-CDHP respectively.}

\emph{Proof.} The security of the data authentication protocol relies on the intractability of \emph{k}-CCAP and inv-CDHP. It can be deduced similarly as the security proof in \cite{CY07}. Due to the page limitation, we omit the proof in detail.

\noindent\textbf{\emph{Theorem 2.}} \emph{The proposed EPDA is unconditionally anonymous.}

\emph{Proof.} Although $V_{ID_{i}}$ is randomly selected in $G_{1}$, there is always a $r'$ satisfying $(h+r')S_{ID_{i}} = V_{ID_{i}}$ for each client $i$ ($i\in\{1, \cdot\cdot\cdot, n\}\setminus\{\hat{a}\}$), which is similar to $(h+r)S_{ID_{\hat{a}}} = V_{ID_{\hat{a}}}$. It is impossible for any adversary to reveal the identity of a client from $V_{ID_{i}}$, thus we can ensure anonymity of the participants in MCS system.

\section{Performance Analysis}

\subsubsection{Computational comparison with other schemes}
We now compare our scheme with other similar schemes in \cite{CY07,ZZW07,Wan10}. We mainly consider the time-consuming operations including the bilinear pairing operation (BP), scalar multiplication in $G_1$ (SM), exponentiation in $G_2$ (EXP) and hash operation (Hash), n is the number of clients. The number of these operations in the selected schemes are shown in Table \ref{tabl:comparison}.

\begin{table*}[htbp]
\scriptsize
\caption{Comparison between existing schemes}\label{tabl:comparison}
\begin{center}
\begin{tabular}{c | c c c c | c c c c}
  \hline \hline
  &\multicolumn{4}{c|}{\raisebox{-0.05cm}{\textbf{Sign}}} & \multicolumn{4}{c}{\raisebox{-0.05cm}{\textbf{Verify}}}\\

  \cline{2-9} \raisebox{0.15cm}[0pt]{\textbf{Algorithm}} & \raisebox{-0.05cm}[0pt]{BP}   & \raisebox{-0.05cm}[0pt]{SM}  & \raisebox{-0.05cm}[0pt]{EXP}   & \raisebox{-0.05cm}[0pt]{Hash}  & \raisebox{-0.05cm}[0pt]{BP}       & \raisebox{-0.05cm}[0pt]{SM} & \raisebox{-0.05cm}[0pt]{EXP} & \raisebox{-0.05cm}[0pt]{Hash}\\

  \hline \hline
  \raisebox{-0.05cm}[0pt]{\quad  \quad CY\cite{CY07} \quad  \quad } & \raisebox{-0.05cm}[0pt]{ \quad 1 \quad } & \raisebox{-0.05cm}[0pt]{\quad 3n-2 \quad } & \raisebox{-0.05cm}[0pt]{\quad 1 \quad } & \raisebox{-0.05cm}[0pt]{\quad n+1  \quad } & \raisebox{-0.05cm}[0pt]{\quad n \quad } & \raisebox{-0.05cm}[0pt]{\quad n \quad } & \raisebox{-0.05cm}[0pt]{\quad 1 \quad } & \raisebox{-0.05cm}[0pt]{\quad 1 \quad }\\

  \hline
  \raisebox{-0.05cm}[0pt]{ZZW\cite{ZZW07}} & \raisebox{-0.05cm}[0pt]{2} & \raisebox{-0.05cm}[0pt]{2n+3} & \raisebox{-0.05cm}[0pt]{n} & \raisebox{-0.05cm}[0pt]{n+1} & \raisebox{-0.05cm}[0pt]{3} & \raisebox{-0.05cm}[0pt]{2n} & \raisebox{-0.05cm}[0pt]{0} & \raisebox{-0.05cm}[0pt]{n+1}\\

  \hline
  \raisebox{-0.05cm}[0pt]{Wang\cite{Wan10}} & \raisebox{-0.05cm}[0pt]{3} & \raisebox{-0.05cm}[0pt]{3n+3} & \raisebox{-0.05cm}[0pt]{0} & \raisebox{-0.05cm}[0pt]{n} & \raisebox{-0.05cm}[0pt]{3} & \raisebox{-0.05cm}[0pt]{2n} & \raisebox{-0.05cm}[0pt]{0} & \raisebox{-0.05cm}[0pt]{n+1}\\

  \hline
  \raisebox{-0.05cm}[0pt]{EPDA} & \raisebox{-0.05cm}[0pt]{1} & \raisebox{-0.05cm}[0pt]{2n-1} & \raisebox{-0.05cm}[0pt]{1} & \raisebox{-0.05cm}[0pt]{1} & \raisebox{-0.05cm}[0pt]{n} &
  \raisebox{-0.05cm}[0pt]{0} & \raisebox{-0.05cm}[0pt]{1} & \raisebox{-0.05cm}[0pt]{1} \\

  \hline \hline

\end{tabular}
\end{center}
\end{table*}

In the signing stage, the proposed EPDA requires only one BP operation that is the most complex operation, while the schemes in \cite{CY07, ZZW07, Wan10} need 1, 2 and 3 such operations respectively. However, in all schemes, the number of BP operation does not increase with the number of users, so it has the least effect on the performance in this stage when the MCS task involves a large number of users. In contrast to BP operation, the computation on the other three types of operations will vary with the number of users. In EPDA, 2n-1 scalar multiplication, 1 exponentiation and 1 Hash operations are involved, so its time consumption in this stage is $2n\times T_{SM}$ approximately, while that in \cite{CY07, ZZW07, Wan10} are about $3n\times T_{SM} + n\times T_{Hash}$, $2n\times T_{SM}+n\times T_{EXP} + n\times T_{Hash}$ and $3n\times T_{SM} + n\times T_{Hash}$ respectively.

In the verification stage, the proposed EPDA requires n bilinear pairing, 1 exponentiation and 1 Hash operations, but no scalar multiplication. So its time consumption in this stage is $n\times T_{BP}$ approximately, while the schemes in \cite{CY07, ZZW07, Wan10} require $n\times T_{BP}+n\times T_{SM}$, $2n\times T_{SM} + n\times T_{Hash}$ and $2n\times T_{SM}+n\times T_{Hash}$ respectively. According to above analysis, EPDA is more efficient with the increasing of the number of users.

\subsubsection{Performance evaluation of EPDA}
In order to evaluate and test the performance of the selected schemes, we first set up a simulation hardware environment to measure the computation overhead of each scheme. The simulation environment is Linux Ubuntu OS over an Intel Pentium G630 2.7 GHz processor and 4096MB memory. The ECC-based function library is pbc-0.5.14. In order to better evaluate the system performance, we assume that there are n users that try to upload their sensing data in a certain time slot. We choose type A curve to complete the simulation. Type A pairings are constructed on the curve $y^2 = x^3 + x$. The algorithms run more efficient and faster over type A curve than other types of curves, especially for the exponentiation computations. So this kind of curve is often used to implement the elliptic curve cryptography.

\begin{figure}
  \centering
  \includegraphics[width=9cm]{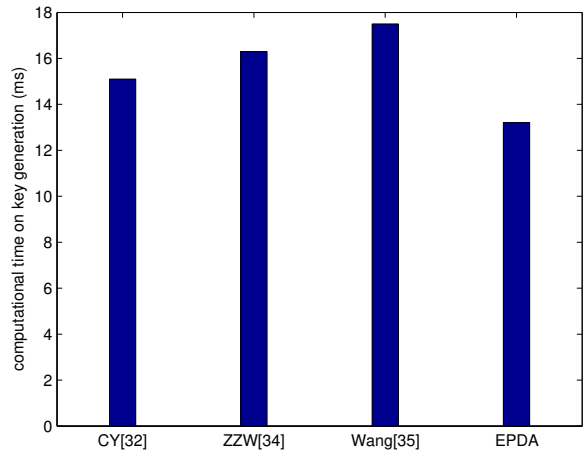}\\
  \caption{Comparison on time consumption of key generation between different schemes}\label{key generation}
\end{figure}

Fig. \ref{key generation} shows that the time overhead on key generation among these schemes is very close, while that in EPDA is least.

\begin{figure*}[t]
\centerline{
\subfigure[time consumption on signing]{\includegraphics[width=6.6cm]{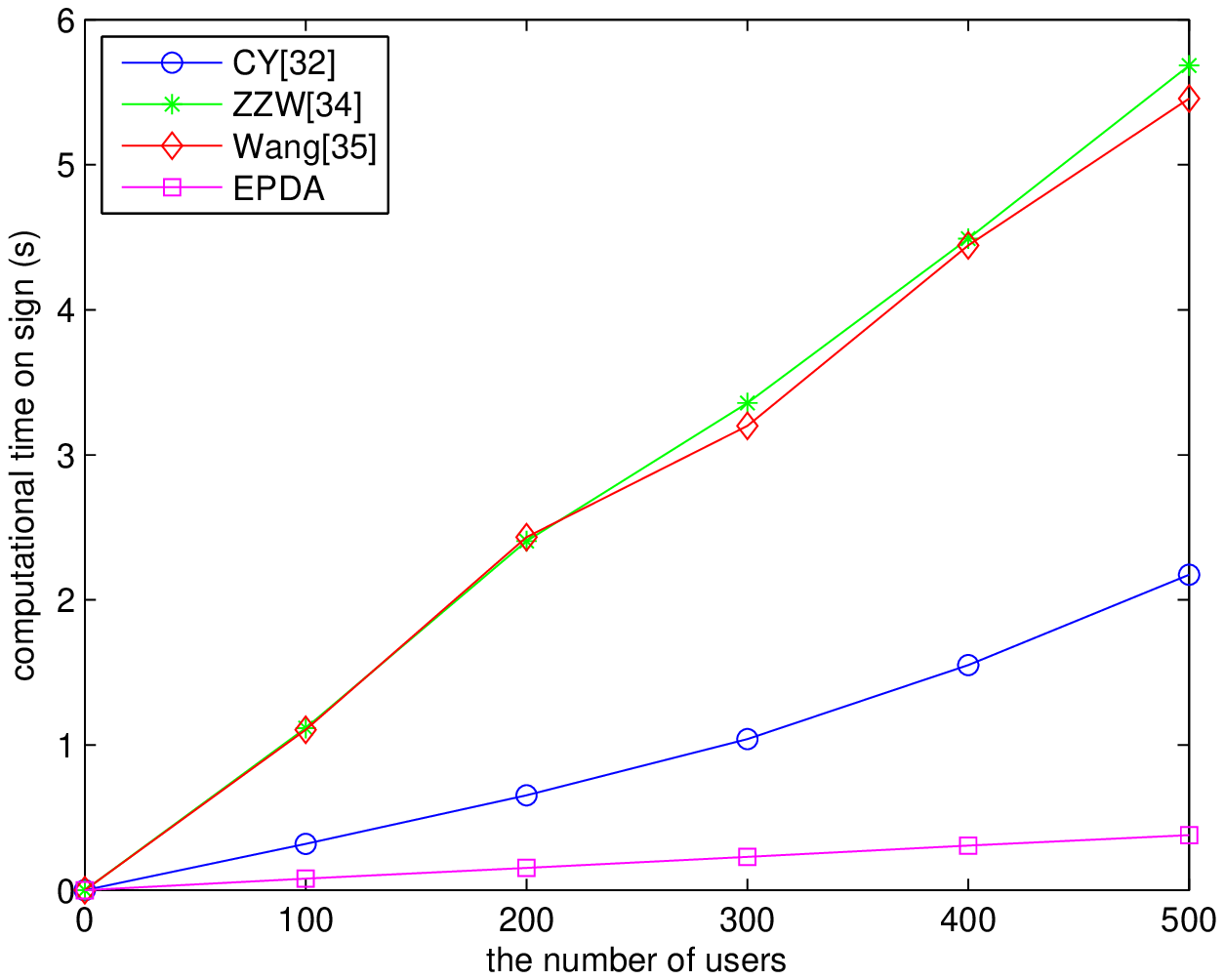}%
\label{sign}} %
%\hfil
\hspace{-7mm}
%\vspace{1cm}
\subfigure[time consumption on verification]{\includegraphics[width=6.6cm]{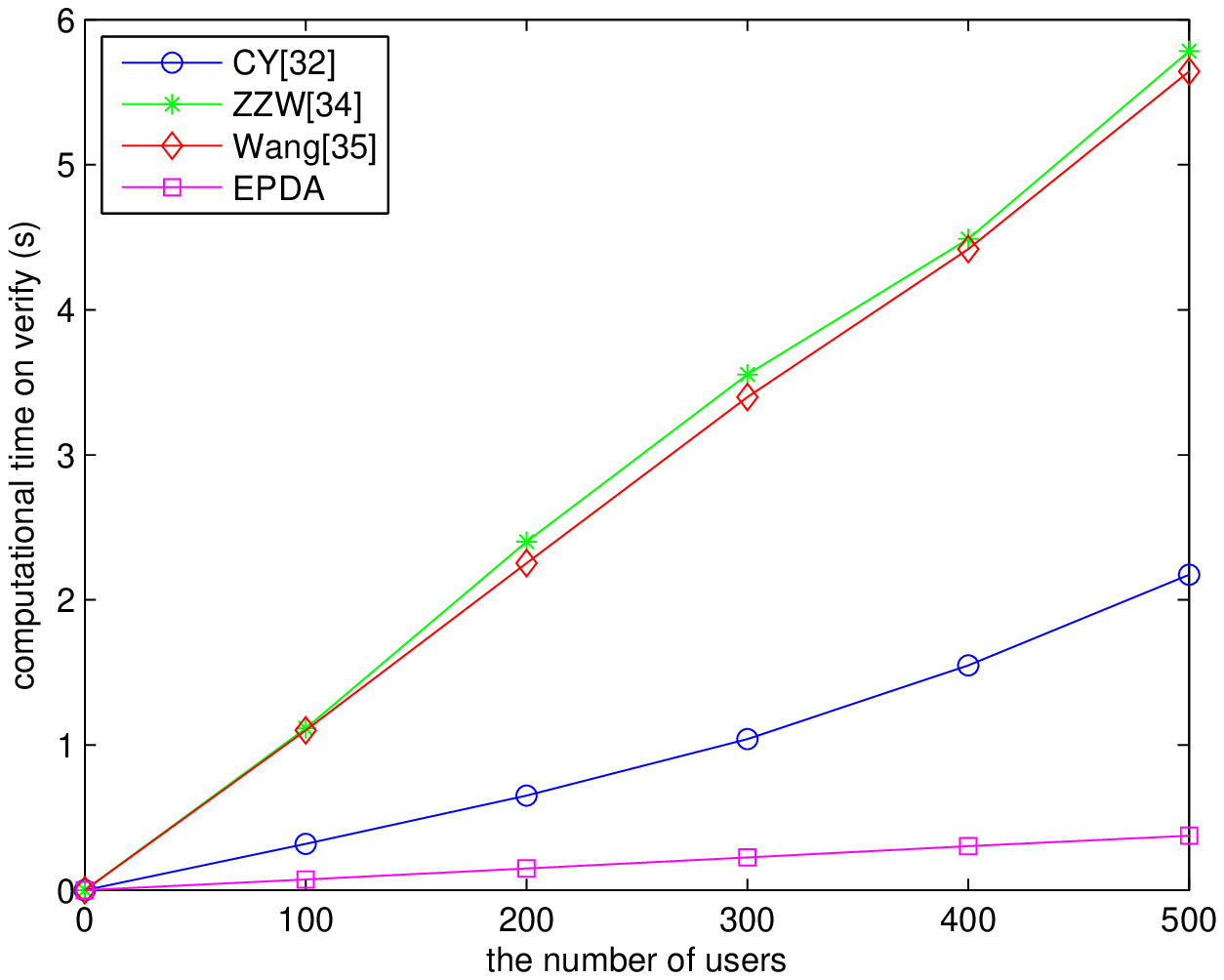}%
\label{verify}} %
%\hfil
\hspace{-7mm}
%%%}
%%%\centerline{
\subfigure[total time consumption]{\includegraphics[width=6.6cm]{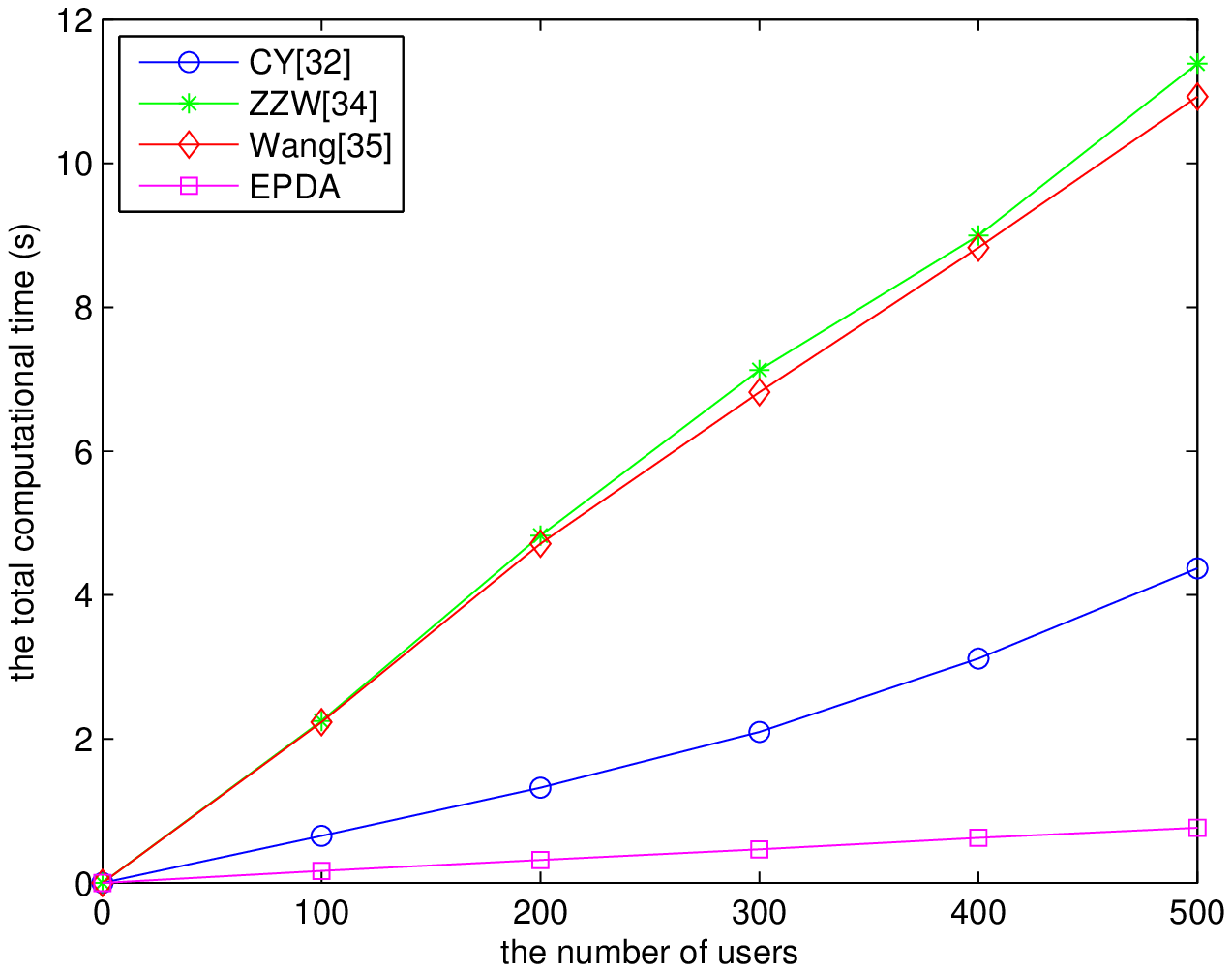}%
\label{total}} %
}
% \captionsetup{skip=-1pt}
%%%\setlength{\abovecaptionskip}{-0.5 mm}
\setlength{\belowcaptionskip}{-2 mm}
\caption{Comparison on time consumption between different schemes}
\label{comparison}
\vspace{-1mm}
\end{figure*}

We repeat the execution of each scheme for 1000 times, and draw up the time consumption by calculating the average value in different stages. Fig. \ref{comparison} shows the comparison of the time consumption on signing, verification, and total time consumption between different schemes with the various number of users, respectively. The scheme in \cite{CY07} requires much less execution time than the schemes in \cite{ZZW07} and \cite{Wan10}, but takes larger amount of running time than EPDA. As the number of users increases, the gap grows rapidly. According to the simulation results, EPDA has the highest efficiency.

\section{Conclusion}
In this paper, we put forward an enhanced privacy-preserving data authentication scheme for MCS scenario, by deploying an improved certificateless ring signature as the cryptographic primitive. The proposed EPDA can be implemented in MCS systems to provide both data authentication and privacy protection with unconditional anonymous verification property. Formal security analysis is also conducted, which lays theoretic foundation to strengthen the soundness of EPDA. The performance comparison between our scheme and the existing schemes shows that the proposed scheme can achieve both low computation complexity and time efficiency. It is an effective solution to the challenges of privacy leak faced by MCS systems.

\section*{Acknowledgements}
This work is supported by Natural Science Basic Research Plan in Shaanxi Province of China (No. 2016JM6057), the 111 Project (B08038) and Collaborative Innovation Center of Information Sensing and Understanding at Xidian University.

\end{document}